\documentclass[a4paper]{article}
\usepackage{graphicx}
\usepackage{fullpage}
\usepackage{hyperref}
\usepackage{float}
\restylefloat{table}

\author{Daan Wilmer}
\title{Empirical Evaluation of Algorithms: Final assignment\\\large{Investigation of ``Enhancing flexibility and robustness in multi-agent task scheduling''}}

\begin{document}
\maketitle

\abstract{
	Wilson et al. propose a measure of flexibility in project scheduling problems and propose several ways of distributing flexibility over tasks without overrunning the deadline.
	These schedules prove quite robust: delays of some tasks do not necessarily lead to delays of subsequent tasks.
	The number of tasks that finish late depends, among others, on the way of distributing flexibility.

	In this paper I study the different flexibility distributions proposed by Wilson et al. and the differences in number of violations (tasks that finish too late).
	I show one factor in the instances that causes differences in the number of violations, as well as two properties of the flexibility distribution that cause them to behave differently.
	Based on these findings, I propose three new flexibility distributions.
	Depending on the nature of the delays, these new flexibility distributions perform as good as or better than the distributions by Wilson et al.
}

\section{Introduction}
Scheduling problems are very common in the real world.
Tasks can often be split up in several activities that have to be performed, with some activities being dependent on the result of other activities.
Equally common in the real world are delays: regularly, unforeseen circumstances extend the needed time for activities beyond their original planning.
The question is: how to cope with these delays?

When the completion time of the whole project is at stake, the answer is simple: start every activity as soon as possible, so the project is finished as soon as possible.
However, sometimes it is not just the project itself, but the individual activities that must be finished in time --- for example because of contracts with other parties.
In this case the number of delayed activities should be minimised.
Wilson, Witteveen, Klos and Huisman studied this problem in \cite{wilson}, using the notion of flexibility: the amount of leeway in the schedule.
They propose adding flexibility to the schedule, distributing it over tasks, to increase robustness.

Wilson et al. test their algorithms using instances from PSPLIB, adding 10\% to the minimal required to be the deadline.
They then randomly extend activities to simulate delays, and test how many violations (delayed activities) there are.
To stabilise the output they run the simulation 150 times.
A more detailed description can be found in their paper\cite{wilson}.

The results they show are quite interesting.
However they give little reasoning about causes for the variations.
The goal of this paper is to look for the underlying causes that drive the differences between algorithms, and differences between results of the same algorithm.

\section{Problem}
In this section I will provide a more detailed picture of the problem area.
I will start with describing the problem that is being solved, followed by the contributions by Wilson et al.
Finally I will describe the problem statement and objective of this paper.

\subsection{Project Scheduling Problem}
The project scheduling problem can be defined as follows: given a set of tasks $T$, each having a length $l_t$, and a set of precedence constraints $C \subset T \times T$, assign each task a start time $s_t$ such that $\forall (t,u) \in C: s_t + l_t \leq s_u$.
In other words: there is a set of tasks with a certain length, and some tasks require others to be finished before they can start.
The tasks that need to be completed for another task to start are called its predecessors, and the tasks that require that task to finish before they can start are called its successors.
To state it in a more formal way: let $P_t$ and $S_t$ be the set of predecessors and successors of $t$, respectively.
Then $P_t = \{u \in T : (u,t) \in C\}$ and $S_t = \{u \in T : (t,u) \in C\}$.

This problem is quite simple and can be solved in polynomial time.
It serves, however, as the base algorithm for some more interesting problems, for example the Resource Constrained Project Scheduling Problem.

Wilson et al. study this problem in a multi-agent setting.
The problem is the same, but tasks are executed by different agents.
These agents want some freedom to decide on when they execute the task, so they need some kind of flexibility.

Wilson et al. add this flexibility by assigning each task not a start time, but a start interval $[a_t,b_t]$ in which the task can be started.
The flexibility of this task is defined as the length of this interval.
The total flexibility of a schedule (an assignment of intervals to tasks) is then defined as the sum of all flexibilities.

To ensure autonomy, the problem needs a small adaptation. 
Considering the start interval $[a_t,b_t]$ of task $t$, the following must hold: $\forall (t,u) \in C: b_t + l_t \leq a_u$.
In other words: the beginning of the start interval of a task cannot be earlier than the latest possible ending time of all its predecessors.
This way both tasks can be started anytime in their own interval, without being hindered by other tasks.

Such a schedule is also quite robust: when one task is delayed, this does not necessarily mean that all following tasks are delayed.
The flexibility can in this case be used to cancel out the delay.
However, sometimes this is not enough and the tasks finishes after it should have been finished; this is called a violation.
The question now becomes: how can flexibility be assigned to tasks in such a way that the number of violations is minimal?

To check the answer to this question, Wilson et al. simulate delays; they randomly pick a portion of the tasks, and these will be delayed by a certain amount.
The number of tasks that is delayed and the amount by which they are delayed are parameters of the simulation.
Of course, they cannot add an infinite amount of flexibility: they impose a deadline at $1.1$ times the shortest possible makespan.
To eliminate fluctuations due to randomisation they run this simulation 150 times.

\subsection{Algorithm and outcomes}
Wilson et al. describe three main methods of assigning flexibility to the tasks: maximal, equalised and weighted.
For the equalised distribution they calculate for each task the maximum possible flexibility, and then minimise the squared flexibility loss.
For the weighted version they assign weights to the different tasks before minimising the flexibility loss, resulting in an increased flexibility for tasks with a higher weight.
They propose three ways of weighting: by the number of direct predecessors, the number of predecessors N steps away (its predecessors, its predecessors' predecessors, etc\dots up to a distance of N), and the total number of predecessors (its predecessors, its predecessors' predecessors, etc\dots until there are no more).
Finally, the maximal distribution maximises the total amount of flexibility in the schedule, and then equalises this flexibility without reducing the total flexibility.

In total this comes down to five distributions: maximal, equalised, weighted by number of predecessors, weighted by number of distance-N predecessors and weighted by all predecessors.
These are all tested, and the results are quite interesting.
The equalised distribution results in the least amount of violations, followed by the weighted distribution by number of predecessors, then the distribution weighted by the number of predecessors at most 5 steps away and finally the distribution weighted by the total number of predecessors. (see also Figure 3 in \cite{wilson}).
An interesting case is the maximal distribution: when the number of delays and the amount by which tasks are delayed are low, this distribution performs the worst of all distributions.
However, when the number of delays or the length of the delays increase, this distribution gains on the others, resulting in less violations than the weighted distributions by total predecessors or distance-5 predecessors when the number of delays or the length of delays become relatively large.

Furthermore, the spread of the number of violations is quite large --- also within results for one distribution.
This can be seen very well in Figure 4 of \cite{wilson}: the variation in the number of violations is quite large, and differs from distribution to distribution.

\subsection{Problem statement}
The results in \cite{wilson} are quite interesting, but a real explanation is missing.
The two questions that I will try to answer in this paper are:
\begin{enumerate}
	\item What properties of the instances cause them to produce more or less violations than other instances and why?
	\item What properties of the flexibility distributions cause them to perform better or worse and why?
\end{enumerate}

\section{Study}
To develop hypotheses about the problem that answer the questions of the problem statement, I performed a study to investigate the algorithm and currently available results.
In this study I looked for correlations between characteristics of the problem and performance in terms of number of violations.

These correlations were calculated using the \texttt{cor()} function in R,\footnote{\url{http://www.r-project.org/}} calculating the correlation coefficient between two data sets.
I chose for this approach over graphs because of the amount of data points: there are 600 instances and four algorithms to compare.
This produces a very dense cloud in which none but the strongest correlations can be seen.

\subsection{Instance properties}
I first studied some instance properties to find correlations to the output.
These instances have many measurable properties, but I narrowed the number of options down to four.
I chose these, based on network descriptiveness --- how much information does the metric give on the network of tasks --- and ease of implementation, to increase time available for research.
\begin{description}
	\item[Average tight width] The average number of simultaneous tasks per time unit when all tasks are scheduled at their earliest start time.
		For each task $a_t = b_t = est_t$, and it is counted for all time units in $[a_t,b_t+l_t)$.
		This measures the amount of parallelism --- the number of tasks that can be scheduled in parallel --- of the instance.
		This metric is similar to the inverse of the $I_2$ metric defined in \cite{measures} for measuring network characteristics.
	\item[Average filled width] The average number of simultaneous tasks per time unit with a filled schedule.
		This schedule is calculated as follows: $a_t = est$, $b_t = min_{(t,u) \in C} \{est_u\} - l_t$.
		Each task is counted in all time units where it could be active: $[a_t,b_t + l_t)$.
		This metric is similar to the previous one, but also takes the flexibility in the system into account.
	\item[Natural flexibility] In the tight schedule, with all tasks scheduled at their earliest start time, not all tasks end at the start time of their earliest successor.
		This extra room I call the natural flexibility of an instance. 
		This calculated as follows: $\displaystyle flex = \sum_{t \in T} min_{(t,u) \in C} \{est_u\} - est_t$.
	\item[Complexity] In Software Engineering, the Cyclomatic or McCabe complexity is a metric that describes the complexity of a function.\footnote{See also \url{https://en.wikipedia.org/wiki/Cyclomatic_complexity}}
		This is done by converting a function into blocks of statements with edges between them if they are reachable, and then taking the difference between the number of edges and the number of nodes.
		Since PSP instances are basically directed graphs with an entry point and exit point, they are similar enough to be able to calculate the cyclomatic complexity for these.
		This metric measures the number of linearly independent paths (paths that cover at least one edge that has not been covered by other paths) through the network of tasks.
\end{description}

\subsubsection{Results}
All correlation data gathered in this part of the study can be found in Appendix \ref{section:results-first-experiment}.
In those numbers, the following three observations can be made:
\begin{enumerate}
	\item It seems that with few and small delays there seems to be no correlation at all.
		When the size of delays increases, the correlations grow stronger, but much more when the number of delays increases.
	\item The average tight width does not seem to be correlated at all with the number of violations.
	\item The complexity seems to be quite strongly correlated to the number of violations.
\end{enumerate}

Of those three observations, the last one shows the strongest correlation between a metric and algorithm performance.
I therefore focus on that observation.
Apparently, instances that have a higher complexity tend to have more violations.
Since the number of tasks is equal in all cases, a higher complexity directly implies that a tasks has more precedence constraints.
This means that, in instances that have a higher complexity, tasks have on average more predecessors and more successors.

One explanation can be found in a causal connection between the number of successors of a tasks and the number of violations.
After all: when a task is delayed more than it has flexibility, the ``excess delay'' propagates to its successors.
Depending on how tight the successors are scheduled to the task (at least one is scheduled tight, otherwise the task itself would have more flexibility), the excess delay propagates to one or more successors.
When a task has on average more successors, excess delays can propagate to more tasks, thereby increasing the number of violations.

Another explanation could be, taken from the paper by Wilson et al.\cite{wilson}, that tasks that have more predecessors are more likely to have at least one predecessor having ``excess delay''.
Since graphs with a higher complexity have more predecessors per task, the probability that a task gets propagated delay is higher, thereby increasing the number of violations.
However, an algorithm based on this reasoning already failed to perform better than --- or even as good as --- a similar algorithm that does not take this reasoning into account\cite{wilson}.
It is therefore less likely that this explanation is correct, although I still take it into consideration.

\subsection{Flexibility distribution}
To gain more insight in the flexibility distribution, I first analysed the distributions.
The graphs can be found in Section \ref{section:flex-distribution}.
In these graphs the bins are chosen $[-2,0],(0,2],(2,3]\dots$, so that the first bin only counts tasks with a flexibility of zero (negative flexibilities are not possible).

The first thing that stands out is the large spike in the first bin, indicating that, on average, between 30\% and 45\% of the tasks get no flexibility at all.
Since this varies almost 15 percentage points, the number of tasks that get no flexibility is the first property of the flexibility distribution to measure and correlate with number of violations.

I add two more properties to measure, in order to further investigate the two explanations from the previous section concerning correlations between complexity and number of violations.
Combined these measures form the following list:
\begin{description}
	\item [Number of zeros] The number of tasks that have exactly $0$ (zero) flexibility.
	\item [Flexibility $\times$ number of predecessors] For each task, multiply the flexibility by the number of predecessors that task has, and then take the sum over all tasks.
		This measures how much the flexibility is concentrated in tasks that have many predecessors.
	\item [Flexibility $\times$ number of successors] For each task, multiply the flexibility by the number of successors that task has, and then take the sum over all tasks.
		This measures how much the flexibility is concentrated in tasks that have many successors.
\end{description}

I investigated these metrics independently of the type of distribution they were generated with.

\subsubsection{Results}
All correlation results can be found in Appendix \ref{section:results-second-experiment}.
Table \ref{table:flex-violations-params} shows the correlations between these metrics and the number of violations, and some very interesting observations can be made.

Depending on the number of delayed tasks and the amount by which they are delayed, the correlation between number of zero-flexibility tasks and number of violations can be quite strong.
The correlation seems the strongest when there are many small delays, and the weakest when there are many large delays.
I suspect this is because of delay propagation: when a task has no flexibility at all, any delay it gets will be propagated to its successors.
This way, when more tasks have no flexibility, delays will be propagated more, resulting in more violations.
However, when the total amount of delay becomes too high, this effect might be negated because tasks get more delay than can be handled by flexibility.
In this case the number of zeros makes little difference: delays will be propagated anyway.
When there are only few delays, the correlation is weaker

The second observation that can be made is that the flexibility multiplied by the number of predecessors is not correlated with the number of violations.
This matches the results in \cite{wilson}, refuting the second explanation --- regarding predecessors --- for the correlation between complexity and number of violations.

The third observation is that the flexibility multiplied by the number of successors has a negative correlation with the number of violations.
This supports the first explanation --- regarding the number of successors --- for the correlation between complexity and number of violations.

\section{Hypotheses}
\label{section:hypotheses}
Concluding from this study, I propose two hypotheses that give an answer to the main question:

\begin{enumerate}
	\item Differences in algorithm performance on different instances are caused by differences in instance complexity, because in instances with higher complexity tasks have more successors so that delays are propagated to more tasks, increasing the number of violations.
	\item Differences between algorithms are caused by two factors:
		\begin{itemize}
			\item The number of tasks that have no flexibility, because when there are more tasks without flexibility delays will be propagated more, causing more violations;
			\item The assignment of flexibility to tasks that have many successors, because when a task has more successors delays will be propagated to more tasks, causing more violations; this can be partially prevented by assigning more flexibility to tasks that have many successors.
		\end{itemize}
		Of these two factors the first one has more impact when delays are relatively small, while the second one has more impact when the total delay is much larger.
\end{enumerate}

The first hypothesis links variance in performance to variance in complexity.
To test this, experiments could be run in which the complexity of the instances is fixed.
All other parameters being equal, this should result in lower variance in the number of violations.

The second hypothesis can be tested by experiments where new flexibility distributions are used.
The two different factors can be tested separately, and might show the best results (least amount of violations, compared to other distributions) at different parameter settings for the delays.
If they are both correct, the first factor will yield the best results with small delays, while experiments regarding the second factor will produce the strongest results when there are many large delays.

If the first factor is correct, a distribution that has less tasks with zero flexibility should perform better.
One such distribution could calculate a minimum flexibility $f_{min}$ that acts as a lower bound for the flexibility per task.
This $f_{min}$ should, of course, be as large as possible.
After every task has this flexibility assigned, the remaining flexibility can be divided according to any of the other distributions.

If the second factor is true, a distribution that assigns more weight to task with many successors will perform better.
One such distribution could be the weighted distribution, where the number of successors per task is used as the weight for that task.

These two distributions do not necessarily exclude eachother.
Therefore, a combination of these two could also be created, by using the lower bound for flexibility for each task and then use the weighted distribution based on number of successors to divide the remaining flexibility.
The performance of this combination distribution will probably be better than that of the equalized distribution, although it might --- for specific parameter settings --- not be as good as the specific distributions

\section{Experiments}
In order to test the hypotheses, I conducted another study and performed several experiments.
For each hypothesis I describe what I did and show the results, concluding on whether the hypotheses are supported by the data.

\subsection{First hypothesis}
To test the first hypothesis I studied the original data.
This is possible because the used instances have only three complexity values: $63$, $100$ and $137$.
I divided the instances by complexity level and took the number of violations for the weighted distribution, based on all predecessors, with 80 \% of tasks delayed by 80 \%.
This settings were chosen because the correlations are the strongest at these settings, from all measurements.

\begin{figure}[ht]
	\centering
	\includegraphics[width=9.5cm]{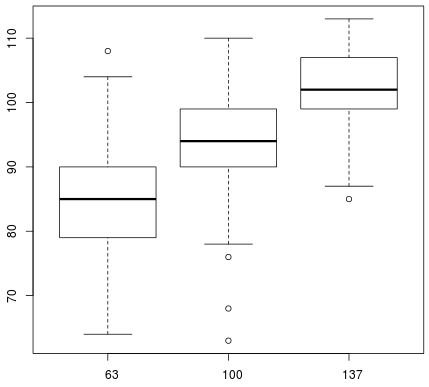}
	\caption{Spread of the number of violations for instances with different complexity values. Weighted distribution, based on all predecessors, with 80 \% of tasks delayed by 80 \%}
	\label{fig:complexity-violations}
\end{figure}

Figure \ref{fig:complexity-violations} shows the spread of the number of violations in instances with different complexity values.
It can quite clearly be seen that the means of the three sets are different, which is verified by a t-test.
What is most important, however, is the variance.
For the combined set the variance is $99.3$; the variance of the subsets with a complexity of $63$, $100$ and $137$ are $66.1$, $49.6$ and $30.0$, respectively.
This supports the statement that differences in instance complexity cause a difference in the number of violations.
There are, however, more factors that cause variance in the number of violations.

\subsection{Second hypothesis}
In order to test the second hypothesis, I experimented with the distribution functions by adapting the original code by Michel Wilson\cite{wilson}.
I added three more flexibility distribution functions:
\begin{description}
	\item[wsucc] Equalise the flexibility over tasks, weighted by number of successors.
	\item[max\_minflex] Maximise the value $f_{min}$ such that every task can get assigned this value as flexibility while the schedule will still be finished before the deadline.
		Then maximise the total flexibility in the schedule while keeping $f_{min}$ as a lower bound for the flexibility of each task.
		Finally equalise the flexibility while ensuring the total flexibility is equal to the maximal flexibility, while keeping $f_{min}$ as lower bound for the flexibility of each task.
	\item[wsucc\_minflex] Maximise $f_{min}$ as described in the previous distribution function, and then equalise the flexibility over the tasks, weighted by number of successors, while keeping $f_{min}$ as a lower bound for the flexibility of each task.
\end{description}

I used two settings for the delay parameters: 80\% of tasks delayed by 5\% and 80\% of tasks delayed by 80\%.
I chose these settings because they yielded the strongest correlations in previous experiments.
The consolidated results of this experiment can be seen in Table \ref{table:performance-new}.

\begin{table}[ht]
	\begin{tabular}{l|rr|rr|}
		 & \multicolumn{2}{c|}{80\% / 5\%} & \multicolumn{2}{c|}{80\% / 80\%}\\
		Flexibility distribution & \#violations & outperforms equalised & \#violations & outperforms equalised\\
		\hline
		Equalised & $31.66$ & n.a. & $70.88$ & n.a. \\
		wsucc & $35.67$ & $22\%$ & $31.66$ & $100\%$ \\
		max\_minflex & $0$ & $100\%$ & $70.63$ & $60\%$ \\
		wsucc\_minflex & $0$ & $100\%$ & $66.17$ & $91\%$ \\
		\hline
	\end{tabular}
	\caption{Performance comparison, averaged over all instances, between the equalised distribution and the new distributions, with 80\% of tasks delayed 5\% (80\%/5\%) and 80\% of tasks delayed 80\% (80\%/80\%)}
	\label{table:performance-new}
\end{table}

It can clearly be seen that, when delays are small, max\_minflex perform best: tasks are never delayed more than their flexibility can compensate.
This makes sense: because the deadline is set at 10\% after the minimum execution time, each task gets assigned at least 10\% of the average task length as flexibility.
Since tasks are delayed 5 \% of their length, violations only occur in tasks that are more than twice as long as the average task and are on the critical path.
Tasks that are not on the critical path can get more flexibility assigned, and therefore need to be even larger to cause a violation when they are delayed.
Apparently such large tasks are not present in the istances.\\
When delays are larger it performs only marginally better than the equalised algorithm.
This matches the weak correlations between the number of zeros and the number of violations.

The distribution that assigns more weight to tasks with many successors, wsucc, shows the opposite effect: when there are many delays, it outperforms the equalised algorithm in all cases.
When delays are smaller, it performs mostly worse than the equalised algorithm.

The distribution that is a combination of the other two introduced distributions, wsucc\_minflex, has also results that are a combination of the results of the other two.
When delays are small, there are no violations; when delays are large, it outperforms the equalised algorithm in most cases.
However, in the latter case, it does not perform as good as the wsucc distribution.

These results conform to the predictions about the second hypothesis stated in Section \ref{section:hypotheses}.
I can therefore conclude that these experiments support the hypothesis.

\section{Conclusion and Future Work}
In this paper I investigated the results by Wilson et al.\cite{wilson}
I raised two questions, which I both answered in this paper.
\begin{enumerate}
	\item \textit{What properties of the instances cause them to produce more or less violations than other instances and why?}
		I found a strong correlation of instances with the number of violations that are created when delays are introduced.
		A causal relation is probable, but has not yet been verified or refuted by further investigation.
	\item \textit{What properties of the flexibility distributions cause them to perform better or worse and why?}
		The study of flexibility distribution showed that it, depending on the delays, the difference between performance are caused by the number of tasks that receive zero flexibility and by the concentration of flexibility in tasks with many successors.
		Further experiments support this conclusion, improving the algorithms from the original paper in two different cases.
\end{enumerate}

The studies and experiments are by no means exhaustive, and therefore the answers to the questions might not be entirely complete.
However, they are quite accurate and help understand the problem and the algorithms.

Possibilities for future work can be found in running more simulations to get a more complete overview of performance with respect to delay parameters.
Besides that, the metrics for the instances and metrics for the flexibility distributions are quite limited in scope.
In further work they could be expanded or improved, to find out whether the answers provided here are the only answers or that there are more factors at work.

Other interesting research can be done investigating other ways of distributing flexibility.
The proposed minflex distribution maximises the lower bound, but it could be interesting to find out if this is not maximised.
Other distributions can include a relative flexibility assignment: assigning a percentage of the task length as flexibility.
For example, when every task gets 10\% flexibility, delays of 10\% can be handled withouth violations.
It should be investigated whether such a distribution is also desirable when delays grow larger.

\appendix
\section{Correlations of instance properties}
\label{section:results-first-experiment}
\begin{table}[H]
	\caption{Correlation of metrics to number of violations for different flexibility distributions, 30\% of tasks delayed 30\%}
	\label{table:metrics-violations-30-30}
	\begin{tabular}{|l|rrrr|}
		\hline
		Metric & Maximal & Equalized & Weighted, direct predecessors & Weighted, all predecessors \\
		\hline
		average tight width & $0.03$ & $-0.02$ & $-0.08$ & $-0.11$ \\
		average filled width & $0.08$ & $-0.08$ & $-0.22$ & $-0.35$ \\
		complexity & $-0.05$ & $0.11$ & $0.26$ & $0.38$ \\
		natural flexibility & $0.06$ & $-0.07$ & $-0.16$ & $-0.27$ \\
		\hline
	\end{tabular}
\end{table}

\begin{table}[H]
	\caption{Correlation of metrics to number of violations for different flexibility distributions, 5\% of tasks delayed 100\%}
	\label{table:metrics-violations-5-100}
	\begin{tabular}{|l|rrrr|}
		\hline
		Metric & Maximal & Equalized & Weighted, direct predecessors & Weighted, all predecessors \\
		\hline
		average tight width & $0.04$ & $0.00$ & $-0.01$ & $0.03$ \\
		average filled width & $-0.10$ & $-0.09$ & $-0.11$ & $-0.13$ \\
		complexity & $0.15$ & $0.14$ & $0.14$ & $0.16$ \\
		natural flexibility & $-0.05$ & $-0.08$ & $-0.09$ & $-0.09$ \\
		\hline
	\end{tabular}
\end{table}

\begin{table}[H]
	\caption{Correlation of metrics to number of violations for different flexibility distributions, 5\% of tasks delayed 5\%}
	\label{table:metrics-violations-5-5}
	\begin{tabular}{|l|rrrr|}
		\hline
		Metric & Maximal & Equalized & Weighted, direct predecessors & Weighted, all predecessors \\
		\hline
		average tight width & $-0.07$ & $-0.03$ & $-0.04$ & $-0.05$ \\
		average filled width & $-0.03$ & $-0.04$ & $-0.06$ & $-0.09$ \\
		complexity & $-0.03$ & $-0.02$ & $0.04$ & $0.09$ \\
		natural flexibility & $0.07$ & $0.01$ & $-0.01$ & $-0.02$ \\
		\hline
	\end{tabular}
\end{table}

\begin{table}[H]
	\caption{Correlation of metrics to number of violations for different flexibility distributions, 80\% of tasks delayed 5\%}
	\label{table:metrics-violations-80-5}
	\begin{tabular}{|l|rrrr|}
		\hline
		Metric & Maximal & Equalized & Weighted, direct predecessors & Weighted, all predecessors \\
		\hline
		average tight width & $0.24$ & $-0.08$ & $-0.24$ & $-0.25$ \\
		average filled width & $0.44$ & $-0.02$ & $-0.41$ & $-0.61$ \\
		complexity & $-0.42$ & $0.05$ & $0.45$ & $0.64$ \\
		natural flexibility & $0.25$ & $0.07$ & $0.20$ & $-0.42$ \\
		\hline
	\end{tabular}
\end{table}

\begin{table}[H]
	\caption{Correlation of metrics to number of violations for different flexibility distributions, 80\% of tasks delayed 80\%}
	\label{table:metrics-violations-80-80}
	\begin{tabular}{|l|rrrr|}
		\hline
		Metric & Maximal & Equalized & Weighted, direct predecessors & Weighted, all predecessors \\
		\hline
		average tight width & $0.04$ & $0.02$ & $-0.06$ & $-0.15$ \\
		average filled width & $-0.34$ & $-0.38$ & $-0.53$ & $-0.66$ \\
		complexity & $0.49$ & $0.52$ & $0.64$ & $0.72$ \\
		natural flexibility & $0.41$ & $-0.44$ & $-0.51$ & $-0.58$ \\
		\hline
	\end{tabular}
\end{table}

\begin{table}[H]
	\caption{Correlation of metrics to flexibility loss for different flexibility distributions}
	\label{table:metrics-flexloss}
	\begin{tabular}{|l|rrr|}
		\hline
		Metric & Equalized & Weighted, direct predecessors & Weighted, all predecessors \\
		\hline
		average tight width & $0.30$ & $0.22$ & $0.05$ \\
		average filled width & $0.50$ & $0.31$ & $-0.07$ \\
		complexity & $-0.60$ & $-0.52$ & $-0.21$ \\
		natural flexibility & $0.25$ & $0.12$ & $-0.12$\\
		\hline
	\end{tabular}
\end{table}

\begin{table}[H]
	\caption{Correlation of metrics to each other}
	\label{table:metrics-metrics}
	\begin{tabular}{|l|rrrr|}
		\hline
		Metric & Average tight width & Average filled width & Complexity & Natural flexibility\\
		\hline
		Average tight width & $1$ & $0.59$ & $-0.56$ & $-0.37$\\
		Average filled width & $0.59$ & $1$ & $-0.83$ & $0.50$\\
		Complexity & $-0.56$ & $-0.83$ & $1$ & $-0.35$\\
		Natural Flexibility & $-0.37$ & $0.50$ & $-0.35$ & $1$\\
		\hline
	\end{tabular}
\end{table}

\section{Correlations of flexibility distribution properties}
\label{section:results-second-experiment}

\begin{table}[H]
	\caption{Correlation of flexibility distribution properties to number of violations for different parameter settings, with $x, y$ meaning $x\%$ of tasks delayed by $y\%$}
	\label{table:flex-violations-params}
	\begin{tabular}{|l|rrrrr|}
		\hline
		Flexibility property & $5, 5$ & $5, 100$ & $30, 30$ & $80, 5$ & $80, 80$ \\
		\hline
		Number of zeros & $0.40$ & $0.28$ & $0.39$ & $0.74$ & $0.09$ \\
		Flexibility $\times$ number of predecessors & $0.02$ & $-0.02$ & $0.00$ & $0.02$ & $-0.13$ \\
		Flexibility $\times$ number of successors & $0.10$ & $-0.01$ & $-0.22$ & $-0.24$ & $-0.48$ \\
		\hline
	\end{tabular}
\end{table}

\begin{table}[H]
	\caption{Average values for characteristics of flexibility distributions for different assignment strategies}
	\label{table:distribution-characteristics}
	\begin{tabular}{|l|rrr|}
		\hline
		Assignment strategy & Number of zeros & Flexibilty $\times$ \#predecessors & Flexibility $\times$ \#successors \\
		\hline
		Maximal  & $53.72$ & $1887$ & $2128.86$\\
		Equalized  & $40.68$ & $1864$ & $2038$\\
		Weighted, direct predecessors  & $42.60$ & $2069$ & $1776$\\
		Weighted, all predecessors  & $46.23$ & $1916$ & $1436$\\
		\hline
	\end{tabular}
\end{table}

\begin{table}[H]
	\caption{Correlation of flexibility distribution characteristics with eachother}
	\label{table:flexibility-flexibility}
	\begin{tabular}{|l|rrr|}
		\hline
		Characteristic & Number of zeros & Flexibilty $\times$ \#predecessors & Flexibility $\times$ \#successors \\
		\hline
		Number of zeros & $1$ & $0.00$ & $0.07$\\
		Flexibilty $\times$ \#predecessors & $0.00$ & $1$ & $0.70$\\
		Flexibility $\times$ \#successors & $0.07$ & $0.70$ & $1$\\
		\hline
	\end{tabular}
\end{table}

\section{Flexibility distribution over tasks}
\label{section:flex-distribution}

\begin{figure}[h!]
	\centering
	\includegraphics[width=9.5cm]{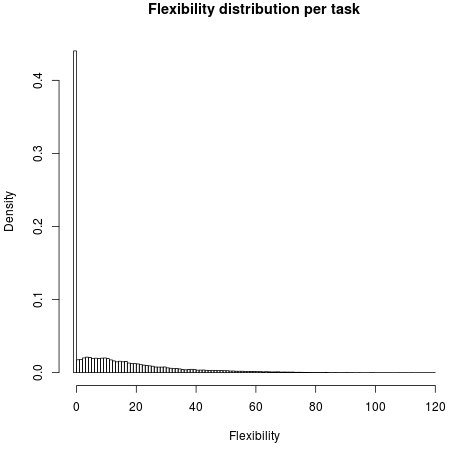}
	\caption{Flexibility distribution per task with maximal distribution, averaged over all instances}
	\label{img:distribution-maximal}
\end{figure}

\begin{figure}[h!]
	\centering
	\includegraphics[width=9.5cm]{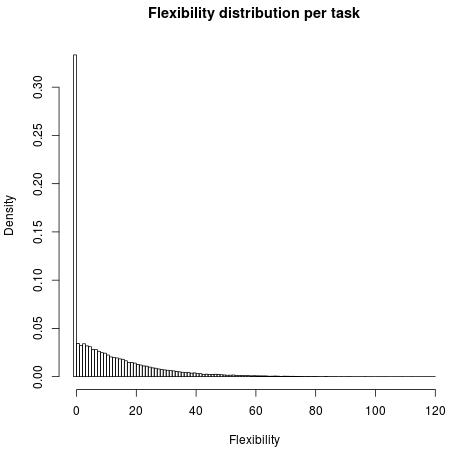}
	\caption{Flexibility distribution per task with equalised distribution, averaged over all instances}
	\label{img:distribution-equal}
\end{figure}

\begin{figure}[h!]
	\centering
	\includegraphics[width=9.5cm]{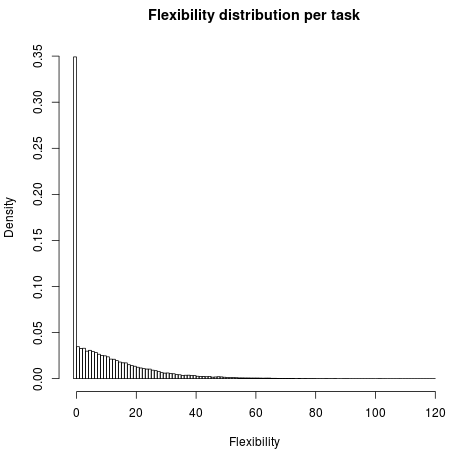}
	\caption{Flexibility distribution per task with weighted distribution by number of predecessors, averaged over all instances}
	\label{img:distribution-wpre}
\end{figure}

\begin{figure}[h!]
	\centering
	\includegraphics[width=9.5cm]{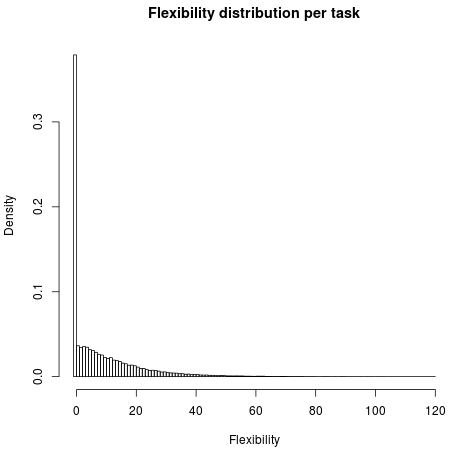}
	\caption{Flexibility distribution per task with weighted distribution by total number of predecessors, averaged over all instances}
	\label{img:distribution-wallpre}
\end{figure}

\bibliographystyle{plain}
\bibliography{bibliography}
\end{document}